\documentclass[12pt]{article}
\usepackage[dvips]{graphicx}
\usepackage{pdproc}

  \makeatletter 
  \def\@cite#1{[#1]} 
  \makeatother    
\begin{document}

\renewcommand{\thefootnote}{\alph{footnote}}

\newlength{\ogltemp}

\def\met{\settowidth{\ogltemp}{\big\slash}%
\hspace{0.5mm}\big\slash\hspace{-1.4\ogltemp}{E_T}%
}
\def\gev{\,\,\mathrm{GeV}\,}
\def\tev{\,\,\mathrm{TeV}\,}
\def\mev{\,\,\mathrm{MeV}\,}
\def\gevc{\,\,\mathrm{GeV}/c\,}
\def\gevm{\,\,\mathrm{GeV}/c^2\,}

%    \vspace*{-3.5cm} 
%  %  \begin{flushright} CDF/PUB/XXXX/PUBLIC/XXXX\\ 
%    \begin{flushright} CDF/PUB/EXOTIC/PUBLIC/7217\\
%    Version 2.0\\ 
%    \today 
%    \end{flushright} 

\vspace*{-1.0cm}

\title{
 Searches for the Supersymmetric Partner of the Bottom Quark
}

\author{ CARSTEN ROTT}

%\address{
%University Department, University Name \\
%Address, City, State ZIP/Zone, Country
%%%%%% You may comment out the e-mail address line below.
%\\ {\rm E-mail: username@abc.def.gh}}

\address{ (for the CDF Collaboration) \\
Purdue University, West Lafayette, Indiana 47907, USA
%%%%% You may comment out the e-mail address line below.  
\\ {\rm E-mail: carott@physics.purdue.edu}}

\abstract{
 We have performed a search for the scalar bottom quark ($\tilde{b}_1$) from 
 gluino ($\tilde{g}$) decays in an R-parity conserving
 SUSY scenario with $m_{\tilde{g}} > m_{\tilde{b}_1}$, 
 by investigating a final state of large missing transverse energy, 
 with three or more jets, and some of them from the hadronization of b-quarks.
 A data sample of $156$~pb$^{-1}$ collected by the Collider
 Detector at Fermilab at a center-of-mass energy of $\sqrt{s}=1.96$~TeV was used.
% by investigating
% a final state of three or more jets, and large missing transverse energy. 
% Events with leptons were vetoed, specific $\Delta\phi$ cuts between jets and missing transverse 
% energy were applied to reduce the QCD multijet background. 
% For the final event selection secondary vertex tags were required. 
 For the final selection, jets containing secondary
 displaced vertices were required.
 This analysis has been performed 'blind', in that the inspection
 of the signal region was only made after the standard model prediction was finalized. 
 Comparing data with SUSY predictions, we can exclude 
 masses of the gluino and sbottom of up to 280 and $240\gevm$ respectively.
%, in an R-parity conserving
% SUSY scenario with $m_{\tilde{g}} > m_{\tilde{b}_1}$.
}

\normalsize\baselineskip=15pt

\section{Introduction}

Despite of its extraordinary success, the Standard Model (SM) 
is incomplete and can be seen as one part of a bigger theory.
%is unsatisfactory and inconsistent with the recent observation of neutrino oscillations, 
%indicating massive neutrinos. 
% The search for physics beyond the SM is 
One attractive extension of the Standard Model is Supersymmetry (SUSY)~\protect\cite{Martin:1997ns}, 
a spacetime
symmetry that relates bosons to fermions and introduces for each SM particle a SUSY partner.
The states $\tilde{q}_L$ and $\tilde{q}_R$ are the partners of the left-handed and 
right-handed quarks, which mix to form mass eigenstates $\tilde{q}_{1,2}$.
For scenarios with large $\tan\beta$ (the ratio of the vacuum expectation values of the 
two Higgs fields), the mixing can be quite substantial in the sbottom sector~\protect\cite{Bartl:1994bu}, so that
the lighter sbottom mass eigenstate (denoted by $\tilde{b}_1$), 
can be significantly lighter than other squarks.

\section{Gluino and Sbottom production at Tevatron}

% Fig:~\ref{cfig1} shows the gluino and sbottom 
% pair production cross-section at $\sqrt{s}=1.96$~TeV
% computed at NLO with PROSPINO~\protect\cite{hep-ph9611232}.
% 
% 
% \begin{figure}[htb]
% \begin{center}
% \includegraphics*[width=6cm]{cfig1.eps}
% \includegraphics*[width=6cm]{cfig2.eps}
% \caption{%
% NLO production cross-section at
% $\sqrt{s}=1.96$ TeV computed with PROSPINO for different
% renormalization scales. Left Figure: direct sbottom pair production. 
% Right Figure: gluino pair production.
% }
% \label{cfig1}
% \end{center}
% \end{figure}
%The next to leading order (NLO) calculation with the PROSPINO program~\protect\cite{hep-ph9611232}
%predicts a cross-section of $2.04$~pb at  $\sqrt{s}=1.96$ TeV for a gluino of 
%mass $240$~GeV/$c^2$, which is large
%compared to the direct production cross section of sbottoms of same mass, which is $0.072$~pb.

The gluino pair production cross-section is expected to be large compared to other SUSY particles.
Next to leading order (NLO) program PROSPINO~\protect\cite{hep-ph9611232} predicts for
a gluino of mass $240\gevm$ a cross-section of $2.04$~pb at  $\sqrt{s}=1.96\tev$, which is large
compared to the direct production cross-section of $0.072$~pb for sbottoms of the same mass.
%The decay of the gluino can only proceed through an on-shell or a virtual squark.
The two body-decays $\tilde{g}\rightarrow q\tilde{q}$, are expected to be the 
dominant gluino decays, if they are allowed, because of the gluino-quark-squark coupling.
% If third generation squarks are lighter than others, 
% it is possible that $\tilde{g}\rightarrow t\tilde{t}_1$ and/or $\tilde{g}\rightarrow b\tilde{b}_1$
% are the only available two-body decay mode(s) for the gluino, in which case they will 
% dominate\footnote{ For a search at the Tevatron the scenario of light sbottoms
% and a gluino decay into sbottom bottom is more attractive. The large top mass
% and restrictions on the stop mass push the needed gluino mass up and make it less favorable
% compared to $\tilde{g}\rightarrow b\tilde{b}_1$.}.

%\section{Search Strategy}
%\section{$\tilde{g}\rightarrow\tilde{b}_1b$, $\tilde{b}_1\rightarrow b\tilde{\chi}_1^0$ Search for Sbottom from gluino decays at CDF II}
\section{Search for Sbottom quarks from Gluino decays at CDF II}

We assume a scenario where the sbottom is lighter than the gluino. 
Further we assume R-parity conservation and the Lightest Supersymmetric 
Particle (LSP), which is stable, to be the lightest neutralino $\tilde{\chi}_1^0$ 
with a mass of $60\gevm$.
%$\tilde{\chi}_1^0$ is the Lightest Supersymmetric Particle (LSP), with a
%mass of $m_{\tilde{\chi}_1^0}=$60~GeV/$c^2$.
%R-parity conservation is assumed which makes it stable.
Gluinos are pair-produced and then decay $100\%$ 
into sbottom bottom ($\tilde{g}\rightarrow\tilde{b}_1b$), followed by the sequential
decay of the sbottom in bottom and lightest neutralino ($\tilde{b}_1\rightarrow b\tilde{\chi}_1^0$).
%$m_{\tilde{\chi}_1^0}$ is assumed to be the Lightest Supersymmetric Particle (LSP), 
%and R-parity is conserved which makes it stable. % and leaves the detector.
Since the neutralinos escape detection, this leaves a signature of four b-jets and 
missing transverse energy ($\met$).
% This analysis was performed assuming we a fixed neutralino
%mass of %$m_{\tilde{\chi}_1^0}=$
%60~GeV/$c^2$.

%We have performed a search for sbottom quarks from gluino decays~\protect\cite{NOTE7136} using the 
%Collider Detector at Fermilab (CDF)~\protect\cite{CDF} at center of
%mass energy  of $\sqrt{s}=1.96$~TeV.

We investigate whether the described scenario~\protect\cite{NOTE7136}
is observable in inclusive three jets events with large $\met$ collected with the 
Collider Detector at Fermilab (CDF), which is described elsewhere~\protect\cite{CDF}.
% The data sample was selected with a trigger that required $\met \equiv |\vec{\met}| \ge 35$~GeV
%and two jet clusters. The missing transverse energy ($\met$) is defined as the negative vector sum
%of the transverse energy in the electomagnetic and hadronic calorimeter.
%Events were selected with $N_{\mbox{jets}} \ge 3$ and 
$\met$ was required to be larger than $35\gev$ and specific clean
up cuts were applied. At the selection stage the data is dominated by QCD multijet,
with the large $\met$ resulting from jet mis-measurements or from semileptonic b-decays
 in which the neutrino escapes detection. In both cases the $\met$ is 
aligned with the %corresponding jet.
mis-measured or b-jet respectively.
The $\met$ in signal originates form the neutralinos from the sequential decay of the gluinos
and is therefore not driven by the jets.
%The $\met$ in signal is uncorrelated with the jets. 
We select events 
in which the $\met$ is not aligned with any of the first three leading jets 
(ordered in $E_T$), which is achieved by computing the opening azimuthal 
angle $\Delta\phi$ between the $\met$ and each of the jets.
By requiring $\Delta\phi(\met,\mbox{1-3jet})>40^\circ$, the standard model background can be effectively
reduced while keeping a large signal acceptance. 
A secondary vertex tagging algorithm is applied to identify b-jets 
and to reduce the background further.
%After this selection the data is divided into four regions, based on the 
Four regions are defined based on the
$\met \in \{(35,50),\ge 50\}\gev$ and the absence or presence of 
high $P_t$ isolated leptons.
We expect our signal to have large $\met$ and no leptons. Hence, we define this
region as our signal regions and the other three regions
serve as control regions.
The two regions with low $\met$ ($35\gev < \met < 50\gev$) serve as control
regions for the QCD multijet background, while the control regions containing
high $P_t$ isolated leptons, are used to check the 
top and W/Z+jets/Diboson background.
%as expected from W and Z boson decays, or top events.
%In this way three control regions and one signal region with events with large $\met$ and 
%without high $P_T$ isolated leptons are defined.
%We can now define four regions, based on the event missing transverse energy 
%$\met \in \{(35,50),\ge 50\}$~GeV and the presents or absence of high $P_t$ isolated leptons.
%The region with large $\met$ and no leptons is our signal region, the other three regions
%serve as control regions.

%The W and Z boson background was estimated using the 
ALPGEN~\protect\cite{alpgen} in combination with the HERWIG~\protect\cite{herwig} event generator
was used to estimate the acceptance of the W and Z boson background.
%event generator in combination with HERWIG~\protect\cite{herwig} event generator. 
The cross-sections at NLO were obtained using the MCFM~\protect\cite{mcfm} program;
The top contributions and the QCD heavy flavor background were predicted using 
the PYTHIA~\protect\cite{pythia} event generator.
%Fake b-tags were estimated using the data.
The fake b-tag contribution, which originates from light flavor jets being 
mis-identified as heavy flavor jets, was estimated using a parameterization
of the fake tag rate obtained from data.

\begin{figure}[htb]
\begin{center}
\includegraphics*[width=5cm]{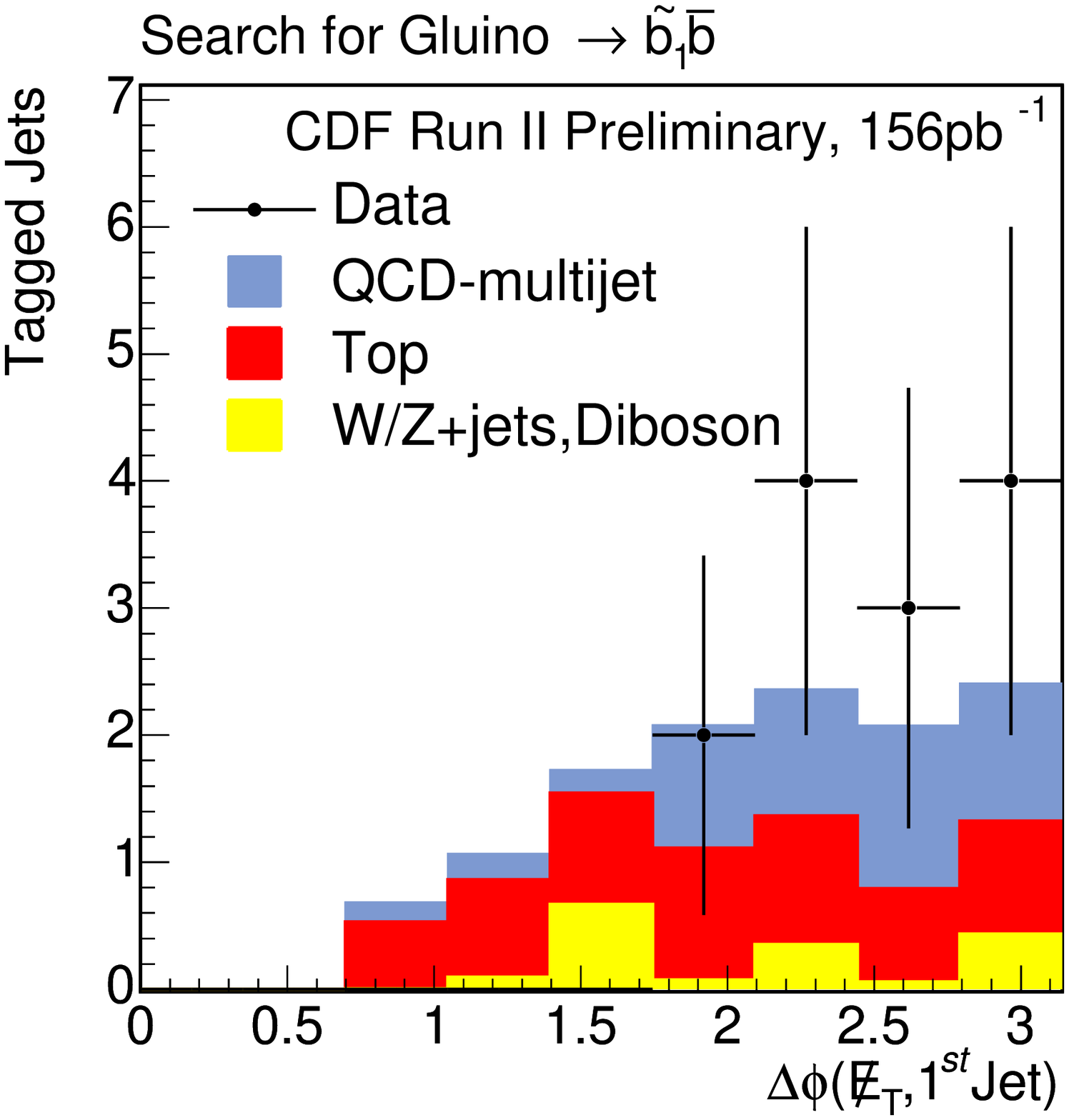}
\includegraphics*[width=5cm]{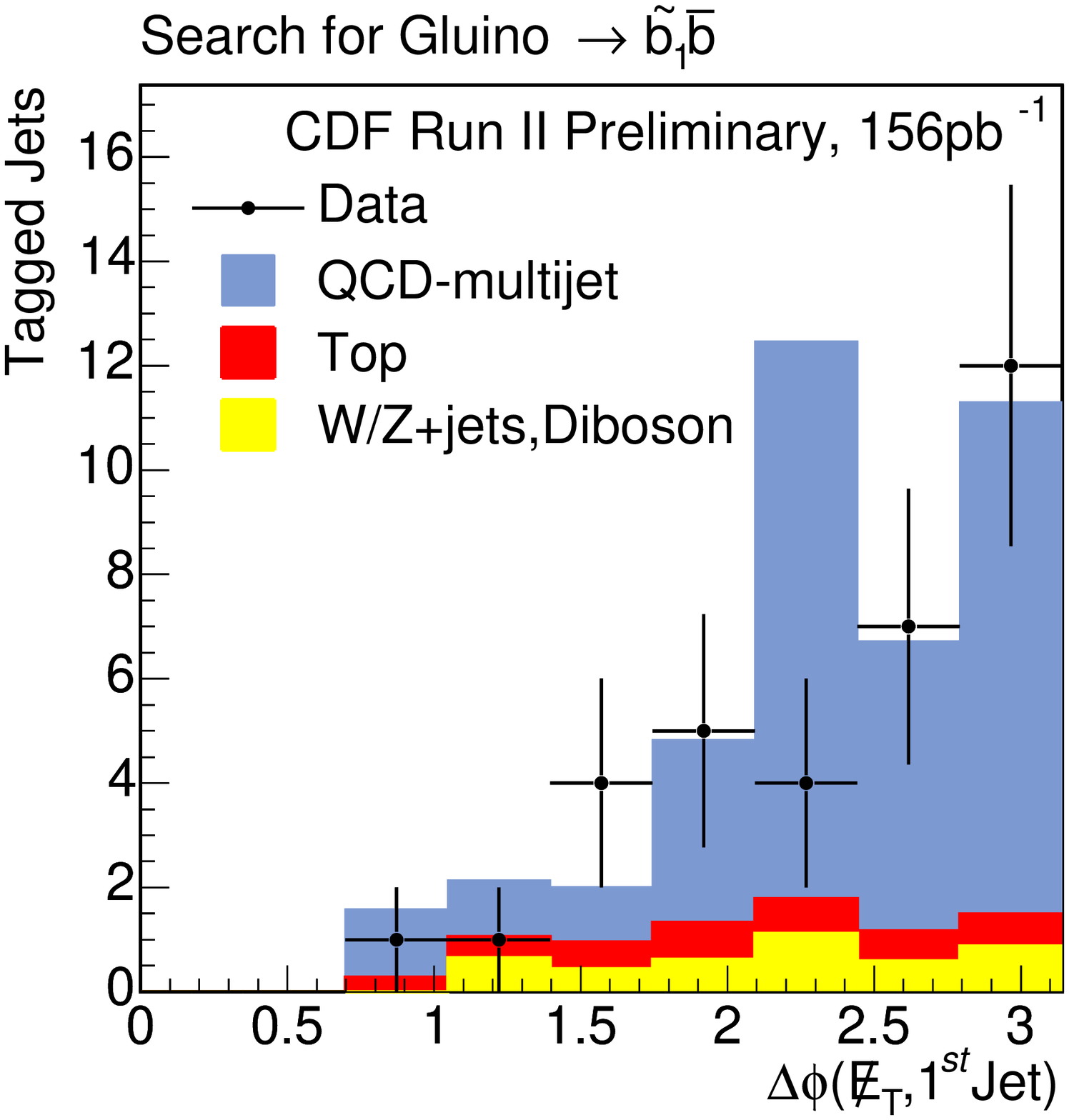}
\includegraphics*[width=5cm]{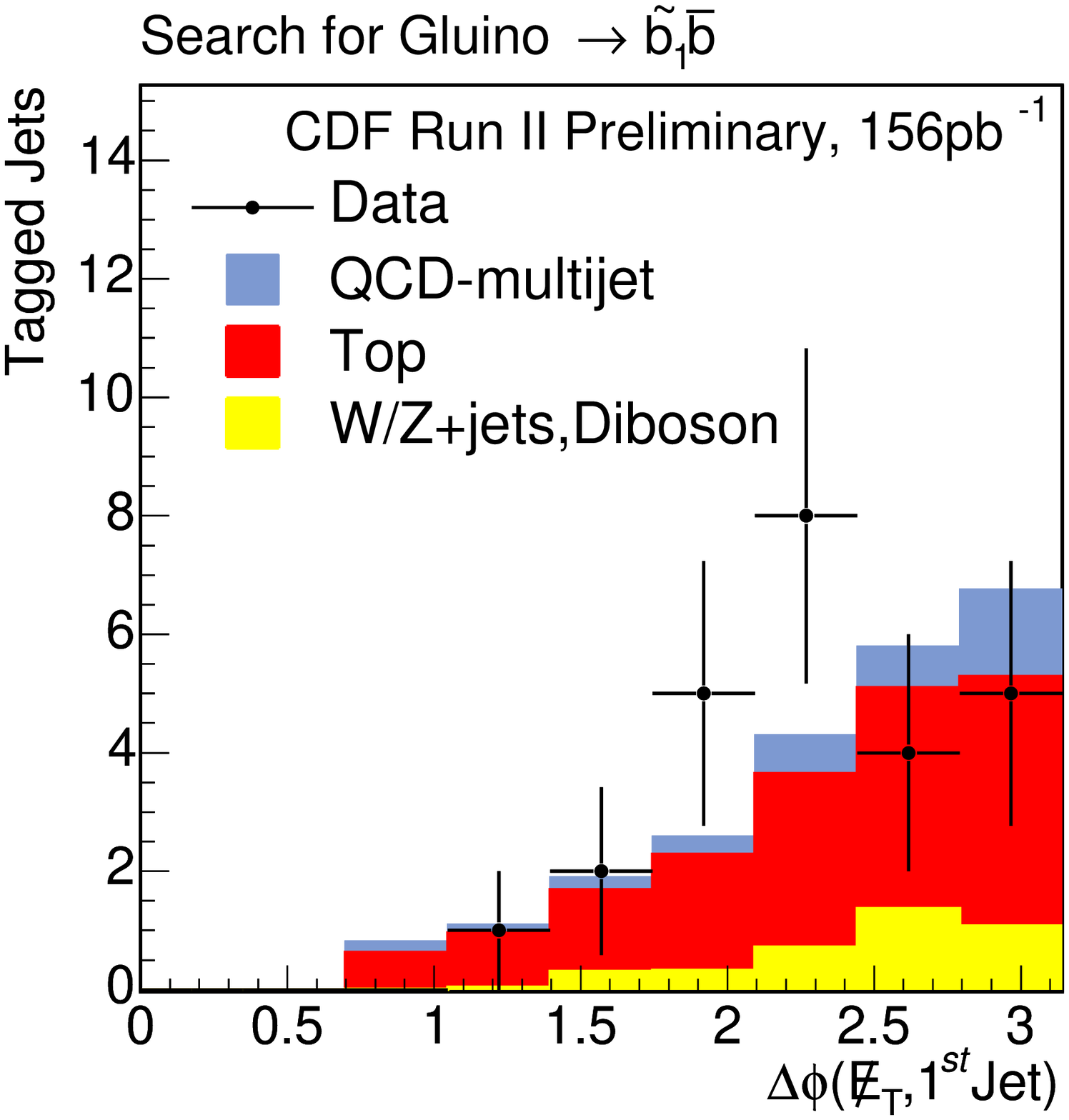}
\caption{%
Comparing observations with prediction in three control regions for 
$\Delta\phi(\met,\mbox{$1^{\mbox{\rm st}}$ jet})$ , where $1^{\mbox{\rm st}}$ jet is the leading jet in the event, if it is tagged.
Left: $35$~GeV$<\met<50$~GeV and high $P_t$ isolated lepton, Middle: $35$~GeV$<\met<50$~GeV and no high $P_t$ isolated leptons 
(QCD multijet dominated control region), Right:$\met>50$~GeV and high $P_t$ isolated lepton (Top dominated control region).}
\label{dphi}
\end{center}
\end{figure}

Various distributions in the control regions have been studied and found to be in
agreement with observations, as an example, Fig.~\ref{dphi} shows the azimuthal opening
angle between the $\met$ and the leading jet $\Delta\phi(\met,\mbox{$1^{\mbox{\rm st}}$ jet})$.

The signal predictions were computed using the ISAJET~\protect\cite{isajet} event
generator with the CTEQ5L %~\protect\cite{CTEQ5L} 
parton distribution functions. 

\begin{figure}[t]
\begin{center}
\includegraphics*[width=6.6cm]{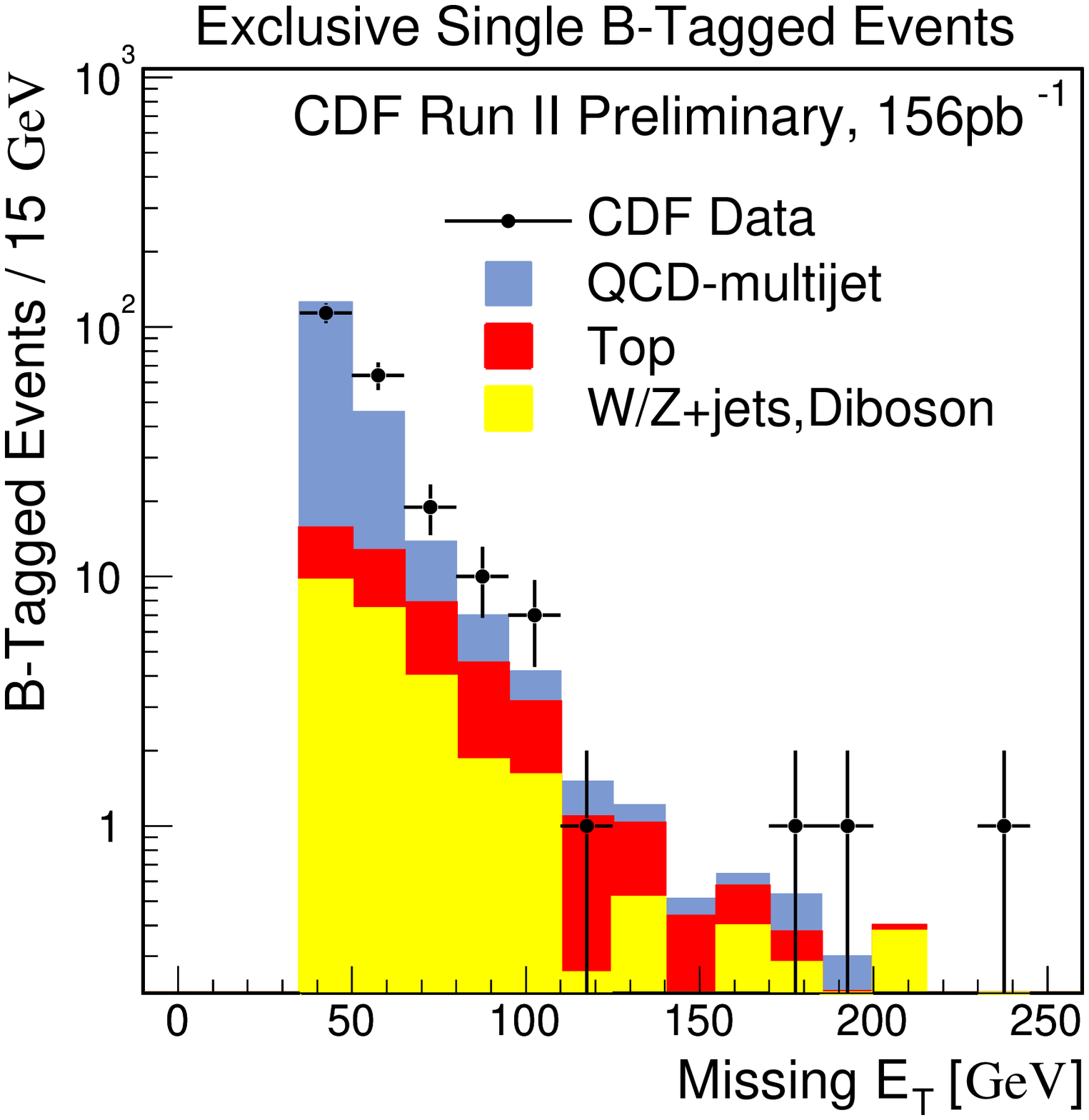}
\includegraphics*[width=6.6cm]{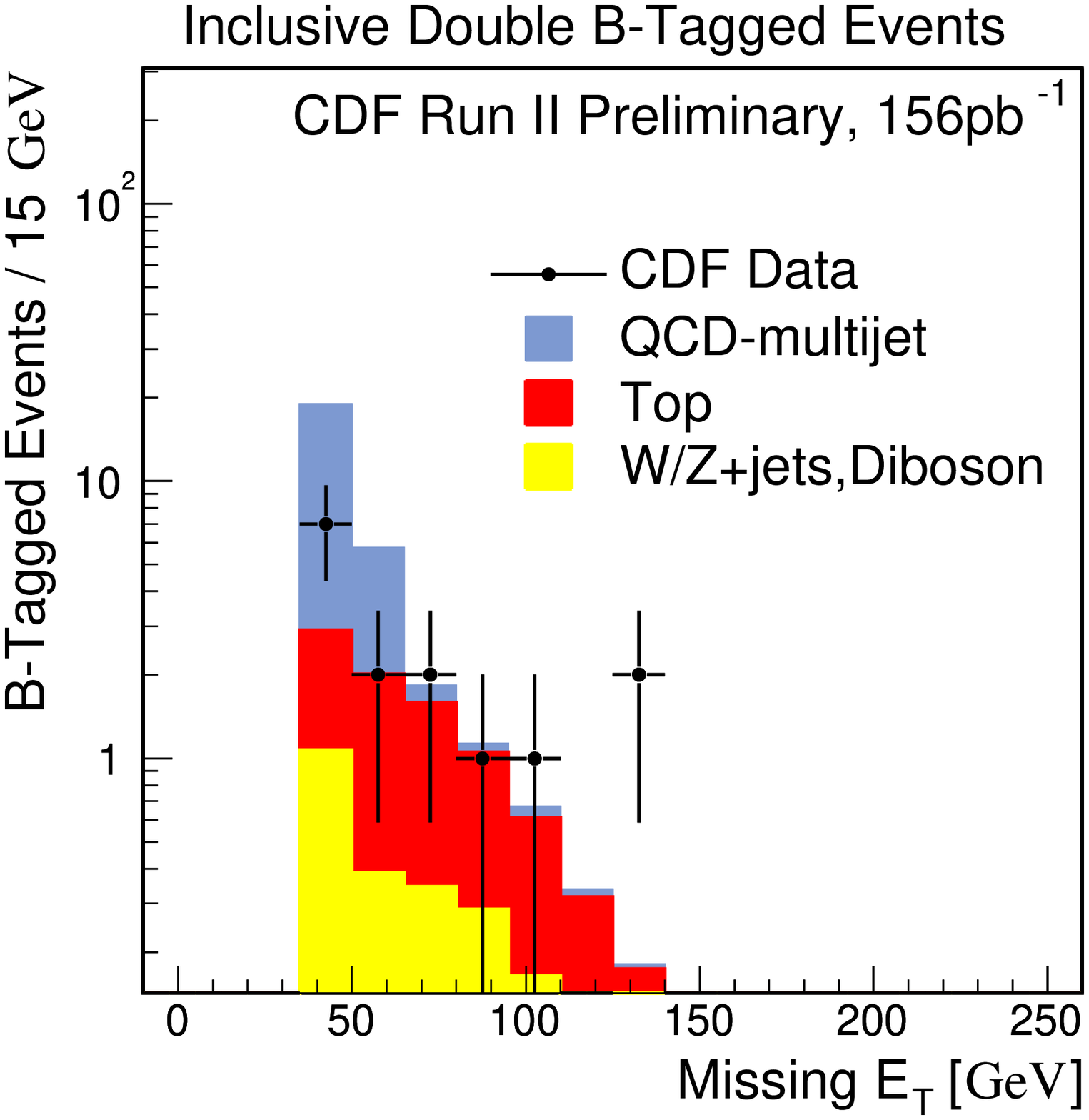}
\caption{$\met$ spectrum after vetoing events with high $P_t$ isolated leptons, for exclusive 
single tagged events (left) and inclusive double tagged events (right).}
\label{fig:spectrums}
\end{center}
%\label{fig:spectrums}
\end{figure}

We perform two analyses using exclusive single b-tagged events and
inclusive double b-tagged events. They serve as an independent cross-check and
in addition the single b-tag analysis is expected to have a better reach 
for nearly mass degenerated gluino-sbottom scenarios. In this case b-jets from the
gluino decays are very soft and less likely to be tagged.
The double tag suppresses the background more effectively by similar signal 
acceptance and is expected to perform better in the other kinematic regions. 
%%The generated events have been passed through the simulation of the CDF detector. 
%The signal event characteristics depend strongly on the mass difference between the gluino
%and the sbottom. For small mass differences, the b-jet from the gluino decays
%are very soft,
%%has low transverse energy due to the reduced phase space, 
%which leads to a reduced acceptance, since events can fail the
%jet requirements and the tagging efficiency is lower for softer
%b-jets.
%%is reduced if one requires multiple jets in the events. 
%%This is because some of the jets
%%will fail the jet requirements and the tagging efficiency is lower for softer
%%b-jets.
%%In the opposite case, the neutralinos from the sbottom decays are
%%produced with a considerable boost and therefore they tend to be back to back,
%%yielding a lower $\met$ in the event.
%Hence, two independent analyses were performed using exclusive single b-tagged events and
%inclusive double b-tagged events. 
Fig.~\ref{fig:spectrums} shows the $\met$ spectrum for both cases.
%for exclusive single tagged events and inclusive
%%double tagged events vetoing leptons. 
%double tagged events 
%fulfilling all signal selection requirements with exception of the $\met$ cut.
The best signal sensitivity was achieved by requiring $\met>80\gev$, which
%The $\met>80$~GeV requirement 
was optimized using signal MC. 
%Signal acceptance systematic uncertainty for the exclusive single tag analysis (16.5\% in total)
Signal acceptance systematic uncertainty for the exclusive single tag analysis (16.5\% in total)
was dominated by jet energy scale (10\%), modelling of initial and final state radiation (7.5\%), 
b-tagging efficiency (7\%), luminosity (6\%), Monte Carlo statistics (3\%),
trigger efficiency (2.5\%), parton distribution functions (2\%), and lepton veto (2\%). 
The uncertainties for the inclusive double tag analysis were very similar, except the b-tagging 
efficiency systematics was increased.

%It was found to be optimal by requiring $\met > 80$~GeV.

The signal region was only analyzed 
after all the background predictions and selection cuts were finalized.
21 exclusive single b-tagged events were observed,
which is in agreement with SM background expectations of $16.4\pm3.7$
events. Requiring inclusive double b-tag we observed $4$ events, where $2.6\pm0.7$
were expected, as summarized in Table:\ref{tab:table1}.

\begin{table}[h]
\begin{center}
\caption{
Number of expected and observed events in the signal region. % with a $\met>80$~GeV.
}
\begin{tabular}{|l|c|c|}
\hline
Process                         &Exclusive Single B-Tag & Inclusive Double B-Tag \\
\hline\hline
W/Z+jets/Diboson & $    5.66 \pm 0.76\rm{(stat)}  \pm  1.72(sys)     $ & $  0.61 \pm 0.21\rm(stat) \pm 0.19(sys)         $ \\
TOP              & $    6.18 \pm 0.12\rm{(stat)}  \pm  1.42(sys)     $ & $  1.84 \pm 0.06\rm(stat) \pm 0.46(sys)         $ \\
QCD multijet     & $    4.57 \pm 1.64\rm{(stat)}  \pm  0.57(sys)     $ & $  0.18 \pm 0.08\rm(stat) \pm 0.05(sys)         $ \\ \hline \hline
Total predicted  & $    16.41 \pm 1.81\rm{(stat)} \pm 3.15(sys)     $ & $  2.63 \pm 0.23\rm(stat) \pm 0.66(sys)         $ \\  \hline
Observed         & $    21                       $ & $  4 $ \\ \hline
\end{tabular}
\label{tab:table1}
\end{center}
\end{table}

\begin{figure}[t]
\begin{center}
\includegraphics*[width=7.5cm]{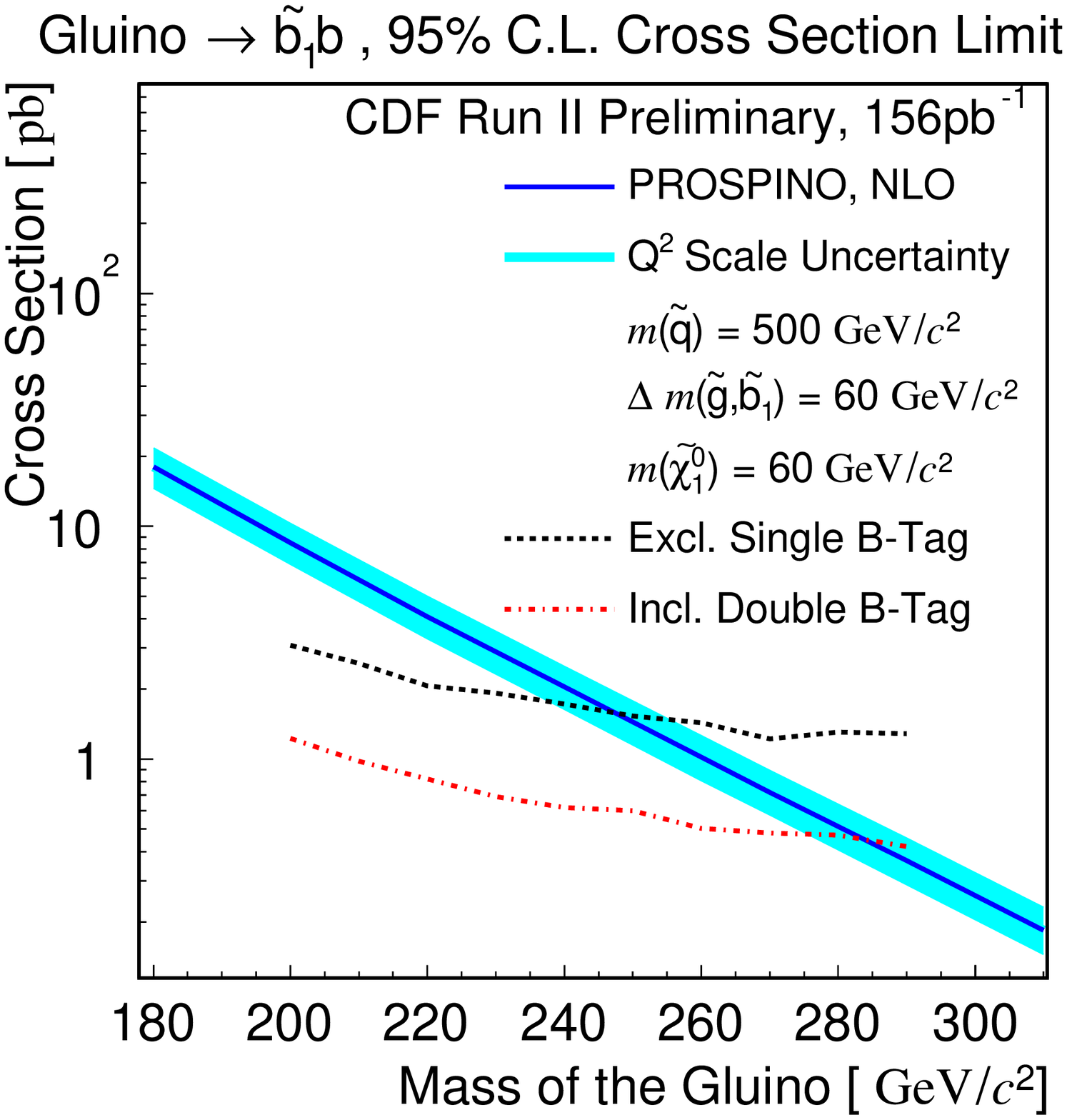}
\includegraphics*[width=7.5cm]{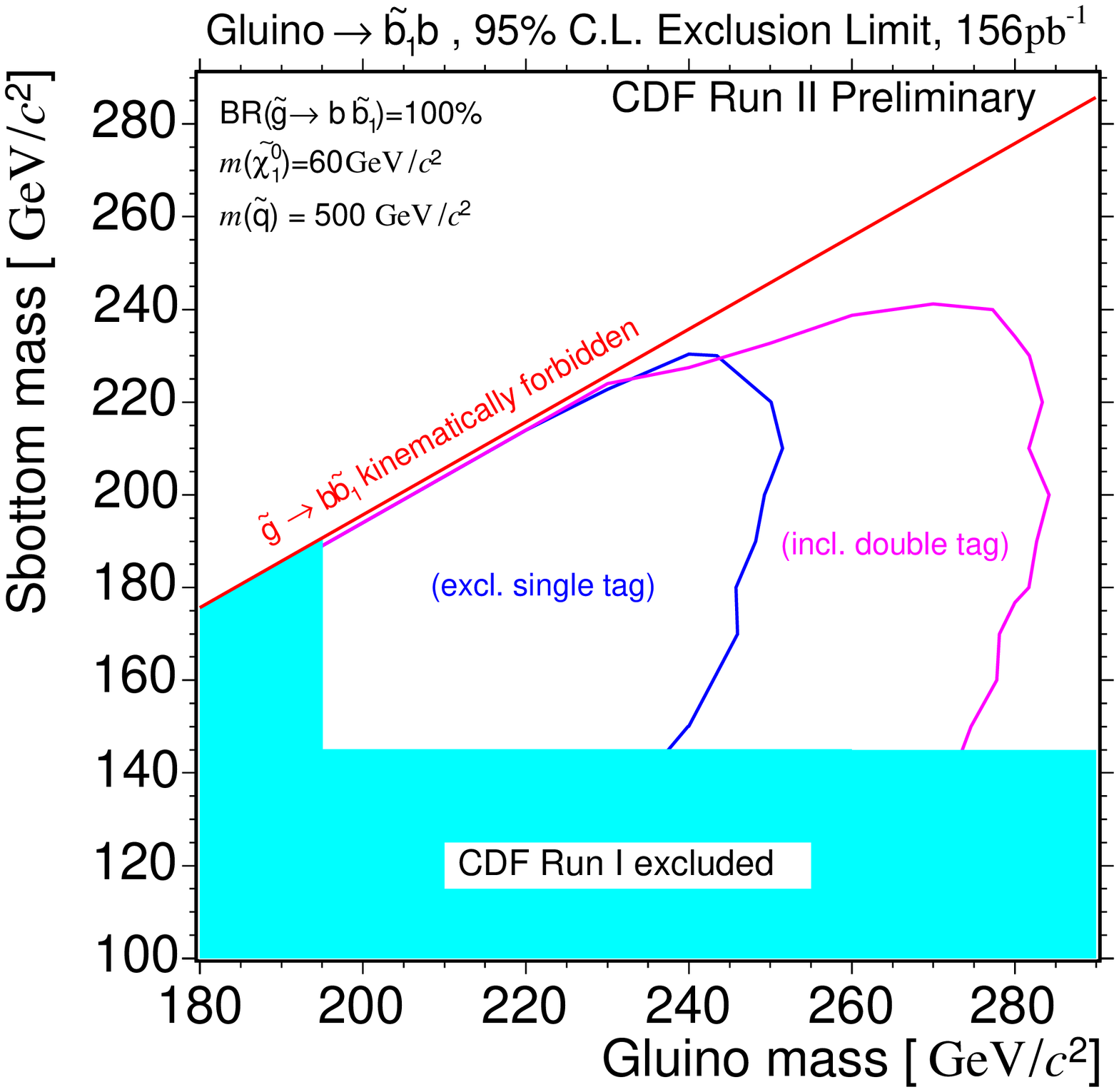}
\caption
{
           Left: 95\% C.L. upper cross-section limit as a function of the gluino mass and
           for a constant mass difference $\Delta m(\tilde{g},\tilde{b}_1) =60\gevm$ between the gluino and sbottom. 
           Right: 95\% C.L. exclusion contours in the $m(\tilde{g})$ and $m(\tilde{b_1})$ plane
           obtained requiring exclusive single b-tagged events and inclusive double b-tagged events.
}
\label{fig:exclusion}
\end{center}
\end{figure}

Since no evidence for gluino pair production with
sequential decay into sbottom-bottom was found, an upper limit cross-sections at
$95\%$ C.L. was computed and an exclusion limit set (see Fig.~\ref{fig:exclusion}) using the
Bayesian likelihood method. 
%The dominant systematics comes form the jet energy scale and tagging efficiency.
%Fig.~\ref{fig:cross_section}  shows the upper limit cross sections for a
%specific gluino mass.

% \begin{figure}[t]
% \begin{center}
% \caption
% {95\% C.L. exclusion contours in the $m(\tilde{g})$ and $m(\tilde{b_1})$ plane.
% }
% \includegraphics*[width=5cm]{cfig8.eps}
% \end{center}
% \label{fig:exclusion}
% \end{figure}

\section{Conclusion}

We have performed a search for gluinos decaying into sbottom bottom at the Tevatron. 
No evidence for this process was found and a $95\%$ C.L. exclusion limit was set on 
the masses of the gluino and sbottom of up to 280 and $240\gevm$ respectively.
%We thank ....

\bibliographystyle{plain}

\begin{thebibliography}{99}
%

%\cite{Martin:1997ns}
\bibitem{Martin:1997ns}
S.~P.~Martin,
``A supersymmetry primer,''
hep-ph/9709356.
%%CITATION = HEP-PH 9709356;%%

%\cite{Bartl:1994bu}
\bibitem{Bartl:1994bu}
A.~Bartl, W.~Majerotto and W.~Porod,
%``Squark and gluino decays for large tan beta,''
Z.\ Phys.\ C {\bf 64}, 499 (1994)
[Erratum-ibid.\ C {\bf 68}, 518 (1995)].
%%CITATION = ZEPYA,C64,499;%%


\bibitem{hep-ph9611232}
W. Beenakker et al.,{\it PROSPINO}, 
hep-ph9611232.

\bibitem{NOTE7136}
CDF/PUB/EXOTIC/PUBLIC/{\bf7136}.

   \bibitem{CDF}
%    F. Abe, et al., Nucl. Instrum. Methods Phys. Res. A {\bf 271},
%    387 (1988);
%    D. Amidei, et al., Nucl. Instum. Methods Phys. Res. A {\bf
%    350}, 73 (1994);
%    F. Abe, et al., Phys. Rev. D {\bf 52}, 4784 (1995);
%    P. Azzi, et al., Nucl. Instrum. Methods Phys. Res. A {\bf
%    360}, 137 (1995);
    The CDF-II Detector Technical Design Report,
    Fermilab-Pub-96/390-E.

    \bibitem{alpgen}
 M.Mangano et al., ALPGEN, %a generator for hard multiparton processes in hadronic collisions, 
hep-ph/0206293.


%\cite{Corcella:2002jc}
\bibitem{herwig}
G.~Corcella {\it et al.},
%``HERWIG 6.5 release note,''
hep-ph/0210213.
%%CITATION = HEP-PH 0210213;%%

%\cite{Campbell:2000bg}
\bibitem{mcfm}
J.~M.~Campbell and R.~K.~Ellis,
%``Radiative corrections to Z b anti-b production,''
Phys.\ Rev.\ D {\bf 62}, 114012 (2000).
%[arXiv:hep-ph/0006304].
%%CITATION = HEP-PH 0006304;%%

%\cite{Sjostrand:2003wg}
\bibitem{pythia}
T.~Sjostrand, L.~Lonnblad, S.~Mrenna and P.~Skands,
``PYTHIA 6.3 physics and manual,''
hep-ph/0308153.
%%CITATION = HEP-PH 0308153;%%

\bibitem{isajet}
H. Baer et al, ISAJET 7.48, %: 
%A Monte Carlo Event Generator for $pp$, $\bar pp$,and $e^+e^-$ Interactions, 
hep-ph/0001086.
    



%    \bibitem{mcfm}
% Calculation of the Zbb and other backgrounds to a ZH signal at the Tevatron.
%J.M. Campbell, R.K. Ellis, Phys. Rev. D62:114012 (2000)%, hep-ph/0006304
%  R.K. Ellis, Sinisa Veseli,Calculation of the Wbb background to a WH signal at the Tevatron, 
%  Phys. Rev. D60:011501 (1999)%, hep-ph/9810489.

%
\end{thebibliography}

\end{document}